\newcommand\two{{(2)}}
\newcommand\one{{(1)}}
\newcommand\zero{{(0)}}
\newcommand\beq{\begin{equation}}
\newcommand\eeq{\end{equation}}
\newcommand\beqa{\begin{eqnarray}}
\newcommand\eeqa{\end{eqnarray}}
\documentclass{elsart}

\usepackage{graphicx,amssymb}
\usepackage{bm}
\usepackage{amsmath}
\journal{Physica A}
\begin{document}
\runauthor{A. Santos}

\begin{frontmatter}



\title{Transport coefficients of $d$-dimensional inelastic Maxwell 
models
} 
\author{Andr\'es Santos}
\address{Departamento de F\'{\i}sica, Universidad de Extremadura,\\
E--06071 Badajoz, Spain}

\date{\today}
\begin{abstract}
Due to the mathematical complexity of the Boltzmann equation for inelastic hard spheres, a kinetic model has recently been proposed whereby the collision rate (which is proportional to the relative velocity for hard spheres) is replaced by an average velocity-independent value. The resulting inelastic Maxwell model has received a large amount of recent interest, especially in connection with the high energy tail of homogeneous states. In this paper the transport coefficients of inelastic Maxwell models in $d$ dimensions are derived by means of the Chapman--Enskog method for unforced systems as well as for systems driven by a Gaussian thermostat and by a white noise thermostat. Comparison with known transport coefficients of inelastic hard spheres shows that their  dependence on inelasticity is captured by the inelastic Maxwell models only in a mild qualitative way. Paradoxically, a much simpler BGK-like model kinetic equation is closer to the results for inelastic hard spheres than the inelastic Maxwell model.
\end{abstract}

\end{frontmatter}

\section{Introduction}
\label{sec1}

As is known, the Boltzmann equation for \textit{elastic} hard spheres 
is in general very complicated  to deal with, so that explicit results are 
usually restricted to small deviations from equilibrium \cite{CC70}. In 
order to explore a wider range of situations, the direct simulation Monte 
Carlo (DSMC) method \cite{B94,AG97} can be used as an efficient tool to 
solve numerically the Boltzmann equation. 
{}From a theoretically oriented point of view,
another fruitful route consists of replacing 
the detailed Boltzmann collision operator by a simpler collision model, e.g. 
the BGK model \cite{BGK54}, that otherwise retains the most relevant 
physical features of the true collision operator \cite{C88}. Several exact solutions of the 
nonlinear BGK model kinetic equation \cite{D90} have proven to agree rather 
well with DSMC results for the Fourier flow \cite{MASG94}, the uniform shear 
flow \cite{OBS89}, the
Couette flow \cite{MG98,MSG00} and the Poiseuille flow \cite{MBG97}.
In a third approach, the mathematical structure of the Boltzmann collision 
operator is retained, but the particles are assumed to interact via the 
repulsive Maxwell potential (inversely proportional to the fourth power 
of the distance) \cite{M67}. For this interaction model, the collision rate 
is independent of the relative velocity of the colliding pair and this 
allows for a number of nice mathematical properties of the collision 
operator \cite{B76,B88,E81}. Since many interesting transport properties 
(both linear and nonlinear), once properly nondimensionalized, are only 
weakly dependent on the interaction potential, exact results derived from 
the Boltzmann equation for elastic Maxwell molecules \cite{SG95} are 
often useful for elastic hard spheres \cite{OBS89} and even Lennard-Jones 
particles \cite{LH87,L88}.
Quoting Ernst and Brito \cite{EB01a,EB01b}, one can say that
``Maxwell molecules are for kinetic theory what harmonic oscillators are 
for quantum mechanics and dumb-bells for polymer physics.'' 

The prototype model for the description of granular media in the regime of 
rapid flow consists of an assembly of (smooth) \textit{inelastic} hard spheres (IHS) 
with a constant coefficient of normal restitution $\alpha$ \cite{C90}.
In the low density limit spatial correlations can be neglected. If, in 
addition, the pre-collision velocities of two particles at contact are 
assumed to be uncorrelated (molecular chaos assumption), the velocity 
distribution function obeys the Boltzmann equation, modified  to account for 
the inelasticity of collisions
\cite{BDS97,vNEB98}. 

Needless to say, all the intricacies of the Boltzmann equation for elastic 
hard spheres are inherited and increased further by the Boltzmann equation 
for IHS, the latter introducing the 
coefficient of normal restitution $0\leq \alpha<1$ in the collision rule. Therefore, it is not 
surprising that the three alternative approaches mentioned above for elastic 
collisions, namely the DSMC method, the model kinetic equation and the 
Maxwell model, have been extended to the case of inelastic collisions as 
well. As for the DSMC method, its extension from the original formulation 
for elastic collisions ($\alpha=1$) \cite{B94,AG97} to $\alpha<1$ is 
straightforward \cite{HS98}.
The generalization of the familiar BGK model kinetic  equation is less 
evident, but a physically meaningful proposal has recently been made \cite{BDS99}. This 
model has proven to yield results in good agreement with DSMC data for the 
simple shear flow problem \cite{BRM97,MGSB99} and for the nonlinear Couette 
flow \cite{TTMGSD01}.
Following the third route, the collision rate of a colliding pair of inelastic 
spheres, which is proportional to the magnitude of the relative velocity, is 
replaced by an \textit{average} constant collision rate \cite{BCG00}. The 
resulting collision operator shares some of the mathematical properties of 
that of elastic Maxwell molecules and so this model is referred to as 
pseudo-Maxwellian model \cite{BCG00} or, as will be done here, inelastic 
Maxwell model (IMM) \cite{EB01a}. The Boltzmann equation for IMM has received a large amount of interest in the last few months
\cite{EB01a,EB01b,BCG00,BNK00,CCG00,C01,KBN01,BC02,BMP01,MP01%
,EB01c,MP02,BNK02a,BNK02b}.

The IMM is worth studying by itself as a toy model to exemplify the 
non-trivial influence of the inelastic character of collisions on the 
physical properties of the system. On the other hand, its practical 
usefulness is strongly tied to its capability of mimicking the relevant 
behaviour of IHS. A common property of the Boltzmann equation for 
IHS and IMM is that both have 
homogeneous solutions exhibiting high energy tails overpopulated with respect 
to the Maxwell--Boltzmann distribution. But this general qualitative 
agreement fails at a  deeper level. More specifically, in the homogeneous 
cooling state the (reduced) velocity distribution function 
$\widetilde{f}(\mathbf{c})$, where $\mathbf{c}$ is the velocity relative to 
the thermal velocity, decays asymptotically as $\ln \widetilde{f}\sim - c$ 
in the case of IHS \cite{EP97,vNE98} and as $\ln \widetilde{f}\sim -\ln c$ 
in the case of IMM \cite{EB01a,EB01b,KBN01,BMP01}. Analogously, in the 
steady homogeneous state driven by a white noise forcing, the asymptotic 
behaviour is $\ln \widetilde{f}\sim - c^{3/2}$ in the case of IHS 
\cite{vNE98}, while $\ln \widetilde{f}\sim - c$ for IMM \cite{EB01c}. 
Of course, these discrepancies in the limit of large velocities do not 
preclude that the the IMM may characterize well the ``bulk'' of the velocity 
distribution of IHS, as measured by low degree moments. In a nonequilibrium 
gas, the physically most relevant moments (apart from the local number 
density $n$, flow velocity $\mathbf{u}$ and granular temperature $T$) are 
those associated with the fluxes of momentum and energy. If the gradients of 
$n$, $\mathbf{u}$ and $T$ are weak enough, the fluxes are linear 
combinations of the gradients, thus defining the transport coefficients 
(e.g. the shear viscosity and the thermal conductivity) as nonlinear functions of the coefficient of 
restitution. These transport coefficients have been obtained by an 
extension of the Chapman--Enskog method \cite{CC70} from the Boltzmann 
equation for IHS \cite{BDKS98,GD99,BC01,GM01}. To the best of my 
knowledge, they have not been derived for IMM. The major aim of this paper is to 
carry out such a derivation and perform a detailed comparison between  the 
transport coefficients for IMM and IHS.

The plan of the paper is as follows. In Sec.\ \ref{sec2} the Boltzmann 
equation for IMM in $d$ dimensions is introduced. The model includes an 
average collision frequency $\omega$ that can  be freely fitted to optimize 
the agreement with IHS. In the absence of any external forcing the energy 
balance equation contains a sink term due to the collisional energy 
dissipation. This term is represented by the cooling rate $\zeta$, that is 
proportional to the collision frequency $\omega$ and to the inelasticity 
parameter $1-\alpha^2$. The sink term can be compensated for by an opposite 
source term representing some sort of external driving. For concreteness, 
two types of  driving are considered: a deterministic force proportional to 
the (peculiar) velocity (Gaussian thermostat) and a stochastic ``kicking'' 
force (white noise thermostat). The corresponding homogeneous solutions are 
analyzed in Sec.\ \ref{sec3}, where the fourth cumulant (or kurtosis) of the velocity distribution is 
exactly obtained, being insensitive to the choice of $\omega$. Comparison 
with the fourth cumulant of IHS shows significant deviations, especially in 
the case of the homogeneous cooling state (which is equivalent to the 
homogeneous steady state driven by the Gaussian thermostat). The 
Chapman--Enskog method is applied in Sec.\ \ref{sec4} to get the transport 
coefficients of IMM in the undriven case, as well as in the presence of the 
Gaussian thermostat and the white noise thermostat. For undriven systems with $d\leq 3$, it is 
found that the  thermal conductivity diverges at $\alpha=(4-d)/3d$ and becomes negative for $\alpha<(4-d)/3d$, irrespective of the choice of $\omega$. A critical comparison with the transport coefficients of IHS is carried out in Sec.\ \ref{sec5}. The free parameter $\omega$ is fixed by the criterion that the cooling rate of IMM be the same as that of IHS (in the local equilibrium approximation). The comparison shows that the IMM retains only the basic qualitative features of the $\alpha$--dependence of the IHS transport coefficients. The best agreement takes place in the case of the white noise thermostat, where the  influence of the inelasticity on the values of the transport coefficients is rather weak.
Quite surprisingly, the transport coefficients predicted by the much simpler BGK-like model \cite{BDS99} are in general much closer to the IHS ones than those obtained from the IMM. The paper ends in Sec.\ \ref{sec6} with some concluding remarks.

\section{Inelastic Maxwell models}
\label{sec2}
The Boltzmann equation for inelastic Maxwell models 
(IMM) \cite{EB01b,BCG00,CCG00} can be obtained from the Boltzmann equation 
for inelastic hard spheres (IHS) by replacing the term  
$|\mathbf{g}\cdot \widehat{\bm{\sigma}}|$ in the collision rate
(where ${\bf g}={\bf v}_1-{\bf v}_2$ is the relative velocity of the 
colliding pair and $\widehat{\bm{\sigma}}$ is the unit vector directed along 
the centres of the two colliding spheres) by an {\em average\/} value proportional to the thermal velocity $v_0=\sqrt{2T/m}$ (where $T$ is 
the granular temperature and $m$ is the mass of a particle). The resulting 
Boltzmann equation is \cite{EB01b}
\beqa
(\partial_t+{\bf v}_1\cdot\nabla+\mathcal{F})f({\bf r},{\bf 
v}_1;t)&=&\frac{\omega({\bf r},t;\alpha)}{n({\bf r},t)\Omega_d}\int 
\d\widehat{\bm{\sigma}}\int \d{\bf 
v}_2
\left(\alpha^{-1}\widehat{b}^{-1}-1\right)
\nonumber\\
&&\times f({\bf r},{\bf v}_1;t)f({\bf r},{\bf 
v}_2;t)\nonumber\\
&\equiv& J[{\bf r},{\bf v}_1;t|f],
\label{1}
\eeqa
where  $n$ is the number density, $\omega(\alpha)\propto n T^{1/2}$ is an 
effective collision frequency, $\Omega_d=2\pi^{d/2}/\Gamma(d/2)$ is the 
total solid angle in $d$ dimensions, $\alpha<1$ is the coefficient of normal 
restitution, and $\widehat{b}$ is the operator transforming pre-collision 
velocities into post-collision ones:
\beq
\widehat{b}{\bf v}_{1,2}={\bf v}_{1,2}\mp\frac{1+\alpha}{2}({\bf 
g}\cdot\widehat{\bm{\sigma}})\widehat{\bm{\sigma}}.
\label{4}
\eeq
Equations (\ref{1}) and (\ref{4}) represent the simplest version of the 
model, since the collision rate is assumed to be independent of the 
relative orientation between the unit vectors $\widehat{\mathbf{g}}$ 
and $\widehat{\bm{\sigma}}$ \cite{EB01b}. In a more realistic version, the 
collision rate has the same dependence on the scalar product 
$\widehat{\mathbf{g}}\cdot \widehat{\bm{\sigma}}$ as in the case of 
hard spheres. The corresponding Boltzmann equation can be proved to be equivalent
to Eq.\ (\ref{1}), except that the operator $\widehat{b}$ must be 
replaced by  \cite{BCG00,CCG00}
\beq
\widehat{b}{\bf v}_{1,2}=\frac{1}{2}\left({\bf 
v}_{1}+\mathbf{v}_2\right)\pm\frac{1-\alpha}{4}\mathbf{g}
\pm\frac{1+\alpha}{4}g\widehat{\bm{\sigma}}.
\label{4bis}
\eeq
Both versions of the model yield similar results in issues as delicate as 
the high energy tails \cite{EB01b,EB01c}. For the sake of 
simplicity, henceforth I will restrict myself to the version of the model 
corresponding to the conventional collision rule (\ref{4}). 

The collision frequency $\omega(\alpha)$ is a free 
parameter of the model. Its detailed $\alpha$--dependence  can be determined 
by optimizing the agreement between the results derived from Eq.~(\ref{1}) 
and those derived from the original Boltzmann equation for IHS.
Of course, the choice of $\omega(\alpha)$ is not unique and may
 depend on the property of interest.

In Eq.~(\ref{1}), $\mathcal{F}$ is an operator representing the action of a 
possible external driving. This operator is assumed to preserve the local 
number and momentum densities, i.e.,
\beq
\int \d\mathbf{v}\, \mathcal{F}f({\bf r},{\bf v};t)
=\int \d\mathbf{v}\, \mathbf{v}\mathcal{F}f({\bf r},{\bf v};t)=0.
\label{b1}
\eeq
On the other hand, in general the external driving gives rise to 
an energy source term 
\beq
\gamma(\mathbf{r},t)=-\frac{m}{d n(\mathbf{r},t)T(\mathbf{r},t)}
\int \d\mathbf{v}\, {V}^2\mathcal{F}f({\bf r},{\bf v};t),
\label{b2}
\eeq
where 
\beq
T(\mathbf{r},t)=\frac{m}{dn(\mathbf{r},t)}\int \d\mathbf{v}\, 
{V}^2f({\bf r},{\bf v};t)
\label{b3}
\eeq
defines the granular temperature and 
$\mathbf{V}=\mathbf{v}-\mathbf{u}$ is the peculiar velocity, 
\beq
\mathbf{u}(\mathbf{r},t)=\frac{m}{n(\mathbf{r},t)}\int \d\mathbf{v}\, 
\mathbf{v}f({\bf r},{\bf v};t)
\label{b4}
\eeq
being the flow velocity.
A possible external driving corresponds to a deterministic nonconservative
force of the form $\frac{1}{2}m\gamma \mathbf{V}$  \cite{EB01c,MS00}. 
It has been widely used to generate nonequilibrium steady states in the context of molecular 
fluids  and can be justified by Gauss's principle of least constraints \cite{EM90,H91}.
The operator $\mathcal{F}$ describing this force is
\beq
\mathcal{F}f(\mathbf{r},\mathbf{v};t)=\frac{1}{2}\gamma(\mathbf{r},t) 
\frac{\partial}{\partial {\bf v}}\cdot \left[\mathbf{V} f(\mathbf{r},\mathbf{v};t)\right].
\label{b6}
\eeq
The most commonly used type of driving for inelastic particles 
consists of a stochastic force in the 
form of Gaussian white 
noise 
\cite{BNK00,CCG00,KBN01,vNE98,WM96,W96,SBCM98,vNETP99,BSSS99,PTvNE01}. 
Its 
associated operator is
\beq
\mathcal{F}f(\mathbf{r},\mathbf{v};t)=
-\frac{\gamma(\mathbf{r},t)T(\mathbf{r},t)}{2m} 
\left(\frac{\partial}{\partial {\bf v}}\right)^2f(\mathbf{r},\mathbf{v};t).
\label{b5}
\eeq

The macroscopic balance equations for the local densities of mass, momentum 
and energy follow directly from Eq.~(\ref{1}) by taking velocity moments:
\beq
D_t n+n\nabla\cdot \mathbf{u}=0,
\label{b7}
\eeq
\beq
D_t\mathbf{u}+\frac{1}{mn}\nabla\cdot\mathsf{P}=\mathbf{0},
\label{b8}
\eeq
\beq
D_tT+\frac{2}{dn}\left(\nabla\cdot\mathbf{q}+\mathsf{P}:\nabla 
\mathbf{u}\right)=-(\zeta-\gamma)T.
\label{b9}
\eeq
In these equations, $D_t\equiv\partial_t+\mathbf{u}\cdot\nabla$ is the 
material time derivative,
\beq
\mathsf{P}(\mathbf{r},t)=m\int\d\mathbf{v}\, \mathbf{V}\mathbf{V}f(\mathbf{r},\mathbf{v};t)
\label{b10}
\eeq
is the pressure tensor,
\beq
\mathbf{q}(\mathbf{r},t)=\frac{m}{2}\int\d\mathbf{v}\, 
V^2 \mathbf{V}f(\mathbf{r},\mathbf{v};t)
\label{b11}
\eeq
is the heat flux, and
\beq
\zeta(\mathbf{r},t)=
-\frac{m}{d n(\mathbf{r},t)T(\mathbf{r},t)}
\int \d\mathbf{v}\, {V}^2J[{\bf r},{\bf v};t|f]
\label{b12}
\eeq
is the cooling rate.
The energy balance equation (\ref{b9}) shows that the existence of a driving 
with the choice $\gamma=\zeta$ compensates for the cooling effect due to the 
inelasticity of collisions. 
In that case, the driving plays the role of a \textit{thermostat} that makes 
the macroscopic balance equations (\ref{b7})--(\ref{b9}) look like those of 
a conventional fluid of elastic particles. On the other hand, the transport 
coefficients entering in the constitutive equations are in general different from those 
of a gas of elastic particles and also depend on the type of thermostat used.
In what follows, I will assume that either $\gamma=0$ (undriven system, 
$\mathcal{F}=0$) or $\gamma=\zeta$ in Eq.\ (\ref{b6}) (Gaussian thermostat)
and Eq.\ (\ref{b5}) (white noise thermostat).

The balance equations (\ref{b7})--(\ref{b9}) are 
generally valid, regardless of the details of the model for inelastic collisions. 
However, the influence of the collision model appears through the 
$\alpha$--dependence of the cooling rate.
In the case of IMM, one can easily prove (cf.\ 
Appendix \ref{appA}) the following relationship between the collision 
frequency $\omega$ and the cooling rate $\zeta$:
\beq
\zeta(\alpha)=\frac{1-\alpha^2}{2d}\omega(\alpha).
\label{10}
\eeq

\section{Homogeneous states\label{sec3}}
Before solving the inhomogeneous equation (\ref{1}) by the Chapman--Enskog 
method, it is necessary to analyze the homogeneous solutions, especially
their deviations with respect to the Maxwell--Boltzmann distribution
as characterized by the fourth cumulant
\beq
a_2\equiv \frac{d}{d+2}\frac{\langle v^4\rangle}{\langle v^2\rangle^2}-1,
\label{new1}
\eeq
where
\beq
\langle v^k\rangle =\frac{1}{n}\int\d \mathbf{v}\, v^k f(\mathbf{v}).
\label{new2}
\eeq
\subsection{Homogeneous cooling state. Gaussian thermostat\label{sec3.1}}
In the absence of any external driving ($\mathcal{F}=0$, $\gamma=0$), 
Eq.~(\ref{b9}) for homogeneous states reduces to $\partial_t T=-\zeta T$.
It is convenient to scale the velocities with respect to the thermal 
velocity $v_0(t)=\sqrt{2T(t)/m}$ and define the scaled quantities
\beq
\widetilde{f}({\bf c},\tau)=n^{-1}v_0^{d}(t)f({\bf v},t),\quad {\bf c}={\bf 
v}/v_0(t),\quad  \d\tau=\omega\d t.
\label{11}
\eeq
Thus, Eq.\ (\ref{1}) reduces to 
\beqa
\left(\partial_\tau+\frac{1-\alpha^2}{4d}\frac{\partial}{\partial{\bf 
c}_1}\cdot {\bf c}_1\right) \widetilde{f}({\bf 
c}_1)&=&\frac{1}{\Omega_d}\int \d\widehat{\bm{\sigma}}\int \d{\bf 
c}_2\left(\alpha^{-1}\widehat{b}^{-1}-1\right)\widetilde{f}({\bf 
c}_1)\widetilde{f}({\bf c}_2)\nonumber\\
&\equiv&\widetilde{J}[\mathbf{c}_1|\widetilde{f}],
\label{12}
\eeqa
where use has been made of Eq.\ (\ref{10}).
It is interesting to remark that Eq.~(\ref{12}) coincides with the 
Boltzmann equation corresponding to a homogeneous \textit{steady} state driven by the 
operator (\ref{b6}) with $\gamma=\zeta$ (Gaussian thermostat). In other 
words, the application of the Gaussian thermostat to a homogeneous
system is equivalent to a 
rescaling of the velocities in the freely cooling case.

The so-called homogeneous cooling 
state (HCS) is characterized by a similarity solution in which all the 
time dependence of $f$ occurs through the scaled velocity $\mathbf{c}$, so 
that it corresponds to a stationary solution of Eq.~(\ref{12}).
Such a solution exhibits an overpopulated high energy tail of 
the form $\widetilde{f}(\mathbf{c})\sim c^{-d-a}$ 
\cite{EB01a,EB01b,KBN01,BMP01}, where, in general, the exponent $a$ depends on the coefficient of restitution.
On the other hand, the ``bulk'' properties of the velocity distribution 
function are associated with low degree moments. Of course, by normalization 
$\langle c^2\rangle=d/2$. Thus, the first non-trivial moment is $\langle 
c^4\rangle$.
Let us multiply both sides of Eq.~(\ref{12}) by $c_1^s$ and integrate 
over ${\bf c}_1$ to get
\beq
\partial_\tau\langle c^s\rangle- (1-\alpha^2)\frac{s}{4d}\langle c^s\rangle 
=-\mu_s\equiv \int \d{\bf c}_1\, c_1^s \widetilde{J}[{\bf 
c}_1|\widetilde{f}].
\label{13}
\eeq 
This hierarchy of moment equations can be solved sequentially and has been analyzed in detail by Ernst and Brito \cite{EB01b}.
For $s=2$ we get the identity $\mu_2=(1-\alpha^2)/4$, which is not but Eq.\ (\ref{10})
in dimensionless form.
The collisional moment $\mu_4$ is evaluated in Appendix \ref{appA} with the 
result
\beqa
\mu_4&=&\frac{1+\alpha}{32d(d+2)}\left\{4\langle 
c^4\rangle\left[3\alpha^2(1-\alpha)-\alpha(17+4d)+3(3+4d)\right]\right.
\nonumber\\
&&\left.-(1+\alpha)d(d+2)(4d-1-6\alpha+3\alpha^2)\right\}.
\label{22}
\eeqa
Inserting this into Eq.\ (\ref{13}) with $s=4$ we get the evolution equation
\beq
\partial_\tau \langle c^4\rangle=-(1+\alpha)^2 
\frac{4d-7+3\alpha(2-\alpha)}{8d(d+2)}\left[\langle 
c^4\rangle-\frac{d(d+2)}{4}
\frac{4d-1-3\alpha(2-\alpha)}{4d-7+3\alpha(2-\alpha)}\right].
\label{b13}
\eeq
In the one-dimensional case ($d=1$), the fourth moment \textit{diverges} as 
$\langle c^4\rangle\sim \exp[\tau (1-\alpha^2)^2/8)]$. This is consistent 
with the fact that in this case the exact stationary solution to 
Eq.~(\ref{12}) is $\widetilde{f}(\mathbf{c})=(2^{3/2}/\pi) (1+2c^2)^{-2}$ 
\cite{EB01a,EB01b,KBN01,BMP01}.
On the other hand, for $d\geq 2$ the moment $\langle c^4\rangle$ relaxes to 
a well defined stationary value (HCS value) whose associated 
 cumulant is
\beq
a_2= \frac{4}{d(d+2)}\langle c^4\rangle -1= 
\frac{6(1-\alpha)^2}{4d-7+3\alpha(2-\alpha)}.
\label{b14}
\eeq

\begin{figure}[tbp]
\includegraphics[width= \columnwidth]{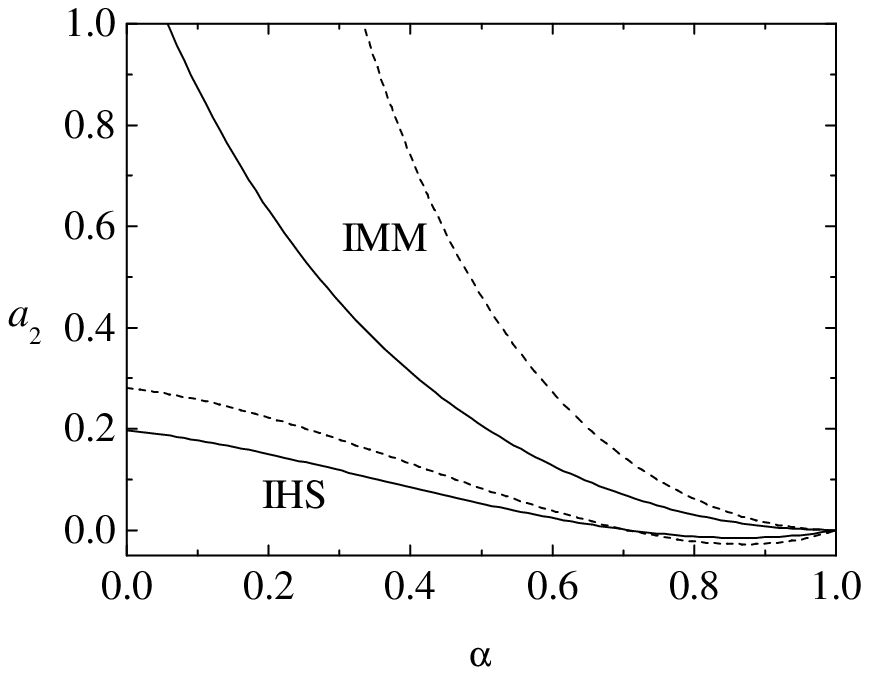}
\caption{Plot of the fourth cumulant $a_2$ in the HCS for $d=3$ (solid 
lines) and $d=2$ (dashed lines).\label{fig1}}
\end{figure}

This expression is exact for the IMM given by Eq.~(\ref{1}). In contrast, 
the cumulant $a_2$ for IHS in the HCS is not known 
exactly. Nevertheless, an excellent estimate is \cite{vNE98}
\beq
a_2^{\text{IHS}}=\frac{16(1-\alpha)(1-2\alpha^2)}{9+24d-\alpha(41-8d)
+30\alpha^2(1-\alpha)}.
\label{b15}
\eeq	
Figure \ref{fig1} compares the result (\ref{b14}) for IMM  with the estimate 
(\ref{b15}) for IHS. It can be observed that the HCS of IMM deviates from 
the Maxwell--Boltzmann distribution (which corresponds to $a_2=0$) much more 
than the HCS of IHS \cite{MSS01}. This is consistent with the fact that the former models 
have a stronger overpopulated high energy tail \cite{EB01a,EB01b,KBN01}, 
$\widetilde{f}(\mathbf{c})\sim c^{-d-a}$, than the latter \cite{vNE98}, 
$\widetilde{f}(\mathbf{c})\sim \e^{-a c}$.

\subsection{White noise thermostat\label{sec3.2}}
\begin{figure}[tbp]
\includegraphics[width= \columnwidth]{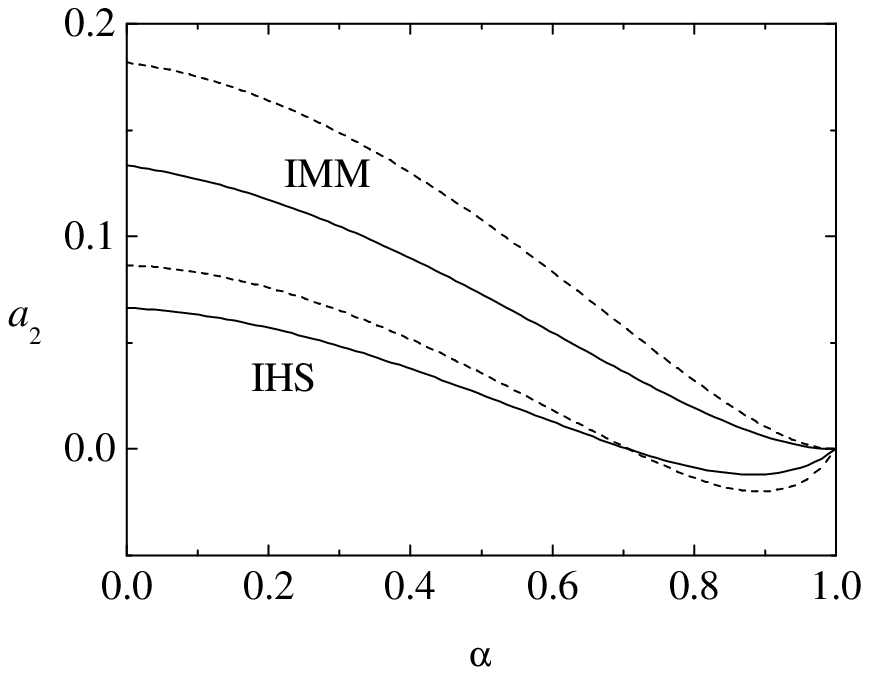}
\caption{Plot of the fourth cumulant $a_2$ in the case of a system heated by 
a white noise thermostat for $d=3$ (solid lines) and $d=2$ (dashed 
lines).\label{fig2}}
\end{figure}

Now we assume that the system is heated with the white noise thermostat 
represented by the operator (\ref{b5}) with $\gamma=\zeta$. Using again the 
scaled quantities (\ref{11}), Eq.~(\ref{1}) becomes
\beq
\left[\partial_\tau-\frac{1-\alpha^2}{8d}\left(\frac{\partial}{\partial{\bf 
c}_1}\right)^2\right] \widetilde{f}({\bf 
c}_1)=\widetilde{J}[\mathbf{c}_1|\widetilde{f}].
\label{24}
\eeq
Taking moments we get
\beq
\partial_\tau \langle 
c^{s}\rangle-\frac{1-\alpha^2}{8d}s(s+d-2)\langle 
c^{s-2}\rangle=-\mu_s.
\label{26}
\eeq
In particular, setting $s=4$,
\beqa
\partial_\tau \langle c^4\rangle&=&-(1+\alpha)
\frac{9+12d-\alpha(17+4d)+3\alpha^2(1-\alpha)}{8d(d+2)}
\nonumber\\
&&\times\left[\langle 
c^4\rangle-\frac{d(d+2)}{4}
\frac{(3-\alpha)\left(4d+5-3\alpha(2+\alpha)\right)}{9+12d-\alpha(17+4d)
+3\alpha^2(1-\alpha)}\right].
\label{b16}
\eeqa
In the case of this thermostat, the moment $\langle c^4\rangle$ relaxes to a steady state value for any 
dimensionality $d$. The corresponding exact expression for the fourth 
cumulant is
\beq
a_2= 
\frac{6(1-\alpha)^2(1+\alpha)}{9+12d-\alpha(17+4d)+3\alpha^2(1-\alpha)}.
\label{27}
\eeq
The cumulant $a_2$ for IHS in the nonequilibrium steady state driven by a 
white noise thermostat can be estimated to be \cite{vNE98}
\beq
a_2^{\text{IHS}}=\frac{16(1-\alpha)(1-2\alpha^2)}{73+56d-3\alpha(35+8d)
+30\alpha^2(1-\alpha)}.
\label{b17}
\eeq	
Figure \ref{fig2} shows that the differences between the values of 
$a_2$ for IMM and
 IHS are less dramatic than in the HCS. The behaviour observed in 
 Fig~\ref{fig2} is in agreement with the  high energy tails 
 $\widetilde{f}(\mathbf{c})\sim \e^{-a c}$ for IMM \cite{EB01c} and
$\widetilde{f}(\mathbf{c})\sim \e^{-a c^{3/2}}$ for 
IHS \cite{vNE98}.

It is interesting to remark that, according to Eqs.~(\ref{b14}) and (\ref{27}), $a_2\propto (1-\alpha)^2$ for IMM in the small inelasticity limit \cite{BCG00,CCG00}, while $a_2\propto (1-\alpha)$ for IHS in the same limit \cite{vNE98}.

\section{Transport coefficients\label{sec4}}
The standard Chapman--Enskog method \cite{CC70} can be generalized to 
inelastic collisions to obtain the dependence of the 
Navier--Stokes transport coefficients on the coefficient of restitution from the Boltzmann equation 
\cite{BDKS98,BC01,GM01} and from the Enskog equation \cite{GD99}. Here the 
method will be applied to the Boltzmann equation (\ref{1}) for IMM.

In the Chapman--Enskog method a factor $\epsilon$ is assigned to every 
gradient operator and  the distribution function is 
represented as a series in this formal ``uniformity'' parameter,
\beq
f=f^\zero+\epsilon f^\one+\epsilon^2 f^{(2)}+\cdots.
\label{c1}
\eeq
Use of this expansion in the definitions of the fluxes (\ref{b10}) and 
(\ref{b11}) and the cooling rate (\ref{b12}) gives the corresponding expansion 
for these quantities. Finally, use of these in the hydrodynamic equations 
(\ref{b7})--(\ref{b9}) leads to an identification of the time derivatives of 
the fields as an expansion in the gradients,
\beq
\partial_t=\partial_t^\zero+\epsilon 
\partial_t^\one+\epsilon^2\partial_t^{(2)}+\cdots.
\label{c2}
\eeq
In particular, the macroscopic balance equations to zeroth order become
\beq
\partial_t^\zero n=0,\quad \partial_t^\zero \mathbf{u}=\mathbf{0}, \quad 
\partial_t^\zero T=-(\zeta-\gamma)T.
\label{c3}
\eeq
Here we have taken into account that in the Boltzmann equation (\ref{1}) the 
effective collision frequency $\omega\propto n T^{1/2}$ is assumed to be a 
functional of $f$ only through the density $n$ and granular temperature $T$. 
Consequently, $\omega^\zero=\omega$, $\omega^\one=\omega^\two=\cdots=0$ and, using Eq.~(\ref{10}), 
$\zeta^\zero=\zeta$, $\zeta^\one=\zeta^\two=\cdots=0$. It must be noticed that, in the case of IHS, $\zeta^\one=0$ and $\zeta^\two$ is small \cite{BDKS98}.

To zeroth order in the gradients the kinetic equation (\ref{1}) reads
\beq
\left(\frac{\zeta-\gamma}{2}\frac{\partial}{\partial \mathbf{V}}\cdot 
\mathbf{V}+\mathcal{F}\right)f^\zero(\mathbf{V})=J[\mathbf{V}|f^\zero],
\label{c4}
\eeq
where use has been made of the properties
\beq
\partial_t^\zero f^\zero(\mathbf{V})=-(\zeta-\gamma)T 
\frac{\partial}{\partial T}f^\zero(\mathbf{V})
=\frac{\zeta-\gamma}{2}\frac{\partial}{\partial \mathbf{V}}\cdot 
\mathbf{V}f^\zero(\mathbf{V}),
\label{c5}
\eeq
the last equality following from the fact that the dependence of $f^\zero$ 
on the temperature is of the form 
$f^\zero(\mathbf{V})=nv_0^{-d}\widetilde{f}^\zero(\mathbf{V}/v_0)$.
In the undriven case ($\mathcal{F}=0$, $\gamma=0$), as well as in the case of 
the Gaussian thermostat (\ref{b6}) with $\gamma=\zeta$, Eq.~(\ref{c4}) is equivalent to the homogeneous equation analyzed in Subsection \ref{sec3.1}. 
Therefore, $f^\zero(\mathbf{V})$ is given by the stationary solution to 
Eq.~(\ref{12}), except that $n\to n(\mathbf{r},t)$ and $T\to 
T(\mathbf{r},t)$ are local quantities and $\mathbf{v}\to 
\mathbf{V}=\mathbf{v}-\mathbf{u}(\mathbf{r},t)$.
Analogously, in the case of the white noise thermostat (\ref{b5}) with 
$\gamma=\zeta$ Eq.~(\ref{c4}) is equivalent to the homogeneous equation analyzed 
in Subsection \ref{sec3.2}.

Since $f^\zero$ is isotropic, it follows that
\beq
\mathsf{P}^\zero=p\mathsf{I},\quad \mathbf{q}^\zero=\mathbf{0},
\label{c6}
\eeq
where $p=nT$ is the hydrostatic pressure and $\mathsf{I}$ is the $d\times d$ 
unit tensor. Therefore, the macroscopic balance equations give
\beq
D_t^\one n=-n\nabla\cdot\mathbf{u},\quad D_t^\one\mathbf{u}=-\frac{\nabla 
p}{mn},\quad D_t^\one T=-\frac{2T}{d}\nabla\cdot\mathbf{u},
\label{c7}
\eeq
where $D_t^\one\equiv \partial_t^\one+\mathbf{u}\cdot \nabla$.
To first order in the gradients Eq.~(\ref{1}) leads to the following 
equation for $f^\one$:
\beq
\left(\partial_t^\zero+\mathcal{L}+\mathcal{F}\right)f^\one(\mathbf{V})=-\left(D_t^\one 
+\mathbf{V}\cdot\nabla\right)f^\zero(\mathbf{V}),
\label{47}
\eeq
where $\mathcal{L}$ is the linearized collision operator
\beqa
\mathcal{L}f^\one(\mathbf{V}_1)&=&-\frac{\omega}{n\Omega_d}\int 
\d\widehat{\bm{\sigma}}\int \d{\bf V}_2
\left(\alpha^{-1}\widehat{b}^{-1}-1\right)\nonumber\\
&&\times
\left[ f^\zero({\bf V}_1)f^\one({\bf V}_2)+f^\zero({\bf V}_2)f^\one({\bf 
V}_1)\right].
\label{c8}
\eeqa
The collisional integrals of $\mathbf{V}\mathbf{V}$ and $V^2\mathbf{V}$ are evaluated in Appendix \ref{appA}.
{}From the linearization of Eqs.~(\ref{AA9}) and (\ref{A3}) we have
\beq
m\int \d\mathbf{V}\,\mathbf{V}\mathbf{V}\mathcal{L}f^\one(\mathbf{V})=\nu 
\mathsf{P}^\one,
\label{c9}
\eeq
\beq
\frac{m}{2}\int 
\d\mathbf{V}\,V^2\mathbf{V}\mathcal{L}f^\one(\mathbf{V})=\left(\frac{d-1}{d} 
\nu+\frac{d+2}{2d}\zeta\right) \mathbf{q}^\one,
\label{c10}
\eeq
where the collision frequency $\nu$ is
\beq
\nu\equiv \omega \frac{(1+\alpha)(d+1-\alpha)}{d(d+2)}.
\label{c11}
\eeq
Using (\ref{c7}), the right-hand side of Eq.~(\ref{47}) can be written as
\beq
-\left(D_t^\one 
+\mathbf{V}\cdot\nabla\right)f^\zero(\mathbf{V})=
{\bf A}(\mathbf{V})\cdot \nabla 
\ln T+{\bf B}(\mathbf{V})\cdot \nabla \ln n+ 
\mathsf{C}(\mathbf{V}):\nabla \mathbf{u},
\label{c12}
\eeq
where
\beq
{\bf A}\equiv\frac{\bf V}{2}\frac{\partial}{\partial {\bf V}}\cdot \left({\bf 
V}f^\zero\right)-\frac{T}{m}\frac{\partial}{\partial {\bf V}}f^\zero,
\label{48}
\eeq
\beq
{\bf B}\equiv-{\bf V} f^\zero-\frac{T}{m}\frac{\partial}{\partial {\bf V}}f^\zero,
\label{49}
\eeq
\beq
C_{ij}\equiv\frac{\partial}{\partial V_i}\left(V_j 
f^\zero\right)-\frac{1}{d}\delta_{ij} \frac{\partial}{\partial {\bf V}}\cdot 
\left({\bf 
V}f^\zero\right).
\label{50}
\eeq

Now we multiply both sides of Eq.\ (\ref{47}) by $m V_i V_j$ and integrate 
over ${\bf V}$. The result is
\beq
(\partial_t^\zero+\nu)P_{ij}^\one+\Pi_{ij}^\one=-p\Delta_{ijkl}\nabla_k u_l,
\label{52}
\eeq
where 
\beq
\Pi_{ij}^\one\equiv m\int\d \mathbf{V}\,V_i V_j \mathcal{F}f^\one(\mathbf{V})
\label{c13}
\eeq
and 
\beq
\Delta_{ijkl}\equiv 
\delta_{ik}\delta_{jl}+\delta_{il}\delta_{jk}-\frac{2}{d} 
\delta_{ij}\delta_{kl}.
\label{c14}
\eeq
Of course, $\Pi_{ij}^\one=0$ in the undriven case. For driven systems,
\beq
\Pi_{ij}^\one=\left\{
\begin{array}{ll}
-\zeta P_{ij}^\one & \text{(Gaussian thermostat)},\\
0& \text{(white noise thermostat)}.
\end{array}
\right.
\label{c15}
\eeq
 The solution to Eq.~(\ref{52}) has the form
\beq
P_{ij}^\one=-\eta \Delta_{ijkl}\nabla_k u_l,
\label{53}
\eeq
where $\eta$ is the shear viscosity.
By dimensional analysis,  $\eta\propto T^{1/2}$. Therefore,
\beq
\partial_t^\zero \mathsf{P}^\one=-\frac{\zeta-\gamma}{2}\mathsf{P}^\one.
\label{c16}
\eeq
Consequently, Eq.~(\ref{52}) yields
\beq
\eta=p
\left\{
\begin{array}{ll}
\left({\nu-\frac{1}{2}\zeta}\right)^{-1} &\text{(undriven system)},\\
\left({\nu-\zeta}\right)^{-1} &\text{(Gaussian thermostat)},\\
{\nu}^{-1} &\text{(white noise thermostat)}.
\end{array}
\right.
\label{54}
\eeq
Except for a possible $\alpha$--dependence of $\nu$ [cf.\ Eq.~(\ref{e3}) below], the result $\eta=p/\nu$ in the case of the white noise thermostat is the same as for elastic particles. This allows us to interpret $\nu^{-1}$ as the effective mean free time associated with momentum transport.
This formal equivalence between the shear viscosity of a fluid of elastic particles and that of a granular fluid driven by a white noise forcing is due to the fact that the latter forcing, while compensating for the inelastic cooling, does not contribute to the rate of change of the stress tensor. The Gaussian thermostat, on the other hand, yields a term $\zeta \mathsf{P}^{(1)}$ and therefore tends to produce a temporal increase in the magnitude of the  stress tensor, thus partially cancelling the dissipative term 
$-\nu\mathsf{P}^{(1)}$. As a consequence, the steady-state shear viscosity is enhanced, $\eta=p/(\nu-\zeta)$. Finally, in the absence of any external driving, the state is unsteady and so the inelastic cooling is responsible for a smaller enhancement of the shear viscosity, $\eta=p/(\nu-\zeta/2)$.
It is worth noting that a structure similar to that of Eq.~(\ref{54}) is also present in the cases of the Boltzmann equation for IHS [cf.\ Eq.~(\ref{B1})] and the BGK-like kinetic model [cf.\ Eq.~(\ref{C3})].

Let us consider next the heat flux. 
Multiplying both sides of Eq.\ (\ref{47}) by $\frac{1}{2}m V^2\mathbf{V}$ 
and integrating over ${\bf V}$ we get
\beq
\left(\partial_t^\zero+\nu' 
\right){\bf q}^\one+\mathbf{Q}^\one=-\frac{d+2}{2}(1+2a_2)\frac{p}{m} \nabla T-\frac{d+2}{2}a_2\frac{T^2}{m}\nabla n ,
\label{60}
\eeq
where 
\beq
\nu'\equiv \frac{d-1}{d}\nu+\frac{d+2}{2d}\zeta=\frac{4(d-1)+(8+d)(1-\alpha)}{4d+4(1-\alpha)}\nu
\label{x1}
\eeq
is an effective collision frequency associated with the thermal conductivity and
\beq
\mathbf{Q}^\one\equiv \frac{m}{2}\int\d \mathbf{V}\,V^2\mathbf{V} 
\mathcal{F}f^\one(\mathbf{V}).
\label{c17}
\eeq
In the undriven case, $\mathbf{Q}^\one=\mathbf{0}$. For the types of driving 
we are considering,
\beq
\mathbf{Q}^\one=\left\{
\begin{array}{ll}
-\frac{3}{2}\zeta \mathbf{q}^\one & \text{(Gaussian thermostat)},\\
\mathbf{0}& \text{(white noise thermostat)}.
\end{array}
\right.
\label{c18}
\eeq
The heat flux has the structure
\beq
{\bf q}^\one=-\lambda \nabla T-\mu \nabla n,
\label{61}
\eeq
where $\lambda$ is the thermal conductivity and $\mu$ is a transport 
coefficient with no counterpart for elastic particles \cite{BDKS98}. 
Dimensional analysis shows that 
$\lambda\propto T^{1/2}$ and $\mu\propto T^{3/2}$. Consequently,
\beqa
\partial_t^\zero {\bf q}^\one &=&\frac{1}{2}(\zeta-\gamma) \lambda\nabla 
T+\frac{3}{2}(\zeta-\gamma) \mu \nabla n+ \lambda\nabla (\zeta-\gamma) 
T\nonumber\\
&=&(\zeta-\gamma)\left[2 \lambda\nabla T+\left(\frac{3}{2}\mu+\lambda 
\frac{T}{n}\right)\nabla n\right],
\label{62}
\eeqa
where in the last step we have taken into account that $\zeta,\gamma\propto 
nT^{1/2}$.
Inserting this equation into Eq.\ (\ref{60}), we can identify the transport 
coefficients as
\beq
\lambda=\frac{p}{m}\frac{d+2}{2}\left(1+2a_2\right)
\left\{
\begin{array}{ll}
\left({\nu'-2\zeta}\right)^{-1} 
&\text{(undriven system)},\\
\left({\nu'-
\frac{3}{2}\zeta}\right)^{-1} 
&\text{(Gaussian thermostat)},\\
{\nu'}^{-1} 
&\text{(white noise thermostat)},
\end{array}
\right.
\label{63}
\eeq
\beq
\mu=\frac{T}{n}\frac{\lambda}{1+2a_2}
\left\{
\begin{array}{ll}
\left({\zeta+a_2
\nu'}\right)\left({\nu'-\frac{3}{2}\zeta}\right)^{-1} 
&\text{(undriven system)},\\
a_2 &\text{(Gaussian 
thermostat)},\\
a_2 &\text{(white noise 
thermostat)}.
\end{array}
\right.
\label{c19}
\eeq
In Eqs.~(\ref{63}) and (\ref{c19}),  the cumulant $a_2$ is given by Eqs.~(\ref{b14}) (undriven system and Gaussian thermostat) and (\ref{27}) (white noise thermostat).

Using Eqs.~(\ref{10}) and (\ref{c11}), it can be seen that the thermal conductivity in the undriven case is $\lambda\propto (\alpha-\alpha_0)^{-1}$, where $\alpha_0=(4-d)/3d$. This implies that the coefficients $\lambda$ and $\mu$ exhibit an 
unphysical behaviour for $d=2$ and $d=3$ since they \textit{diverge} at $\alpha=\alpha_0$ and become negative for 
$0\leq \alpha<\alpha_0$. This singular behaviour is absent in the shear viscosity $\eta$ or in $\lambda$ and $\mu$ for thermostatted states because $\nu>\zeta$ for all $\alpha$ and $d$.

In order to have the full $\alpha$--dependence of the transport coefficients 
we need to fix the free parameter $\omega(\alpha)$. This point will be 
addressed in Section \ref{sec5}. On the other hand, the ratios between 
transport coefficients are independent of the criterion to choose 
$\omega(\alpha)$. Let us define a generalized Prandtl number (for $d\neq 1$)
\beq
R_\eta(\alpha)=\frac{\eta(\alpha)/\eta_0}{\lambda(\alpha)/\lambda_0},
\label{c20}
\eeq
where $\eta_0$ and $\lambda_0$ are the shear viscosity and thermal conductivity, 
respectively, in the elastic limit ($\alpha=1$).
{}From Eqs.~(\ref{54}) and (\ref{63}) we have
\beq
R_\eta(\alpha)=\frac{1}{1+2a_2}
\left\{
\begin{array}{ll}
\frac{d(d-4+3d\alpha)}{(d-1)[3d+2+(d-2)\alpha]} 
&\text{(undriven system)},\\
1&\text{(Gaussian thermostat)},\\
\frac{d[5d+4-(d+8)\alpha]}{4(d-1)(d+1-\alpha)} &\text{(white 
noise thermostat)}.
\end{array}
\right.
\label{c21}
\eeq
Analogously, we can define the ratio
\beq
R_\mu(\alpha)=\frac{n\mu(\alpha)}{T\lambda(\alpha)}.
\label{c22}
\eeq
Thus,
\beq
R_\mu(\alpha)=\frac{1}{1+2a_2}
\left\{
\begin{array}{ll}
 \frac{d+2}{d-1}\frac{1-\alpha}{1+\alpha}+\frac{5d+4-(d+8)\alpha}{2(d-1)(1+\alpha)}a_2
&\text{(undriven system)},\\
a_2&\text{(Gaussian thermostat)},\\
a_2 &\text{(white 
noise thermostat)}.
\end{array}
\right.
\label{68}
\eeq

\section{Comparison with the transport coefficients of inelastic hard 
spheres\label{sec5}}
The transport coefficients of IHS described by the Boltzmann equation
have been derived both for undriven 
\cite{BDKS98,GD99,BC01} and thermostatted \cite{GM01} systems in 
the first Sonine approximation. For the sake of completeness, the expressions of the transport
coefficients of IHS are listed in Appendix \ref{appB}.

Figures \ref{fig3} and \ref{fig4} compare the ratios $R_\eta(\alpha)$, Eq.\ (\ref{c20}),
and $R_\mu(\alpha)$, Eq.\ (\ref{c22}), for IMM and IHS in the 
three-dimensional case.
It is observed that, in general, the results for IMM describe qualitatively the 
$\alpha$--dependence of $R_\eta$
and $R_\mu$ for IHS.
Thus, as the inelasticity increases, the generalized Prandtl number $R_\eta$ 
decreases  in the absence of external forcing and increases in the case of the 
white noise thermostat. With the Gaussian thermostat, however, the discrepancies are important:
$R_\eta$ increases with the inelasticity for IHS and decreases for IMM.
 As for the ratio $R_\mu$, which measures the 
new transport coefficient $\mu$ relative to the thermal conductivity, it 
rapidly increases in the unforced case,  while it is very small in the 
driven cases.
At a quantitative level,  the IMM results exhibit important 
deviations from the IHS ones, especially in the undriven case, where 
$R_\eta$ for IMM becomes negative when $\alpha<\alpha_0=\frac{1}{9}$ and 
$R_\mu$ grows too rapidly.
The situation in which the IMM ratios are the closest to the IHS ones corresponds to 
the system heated by a white noise thermostat.
\begin{figure}[tbp]
\includegraphics[width= \columnwidth]{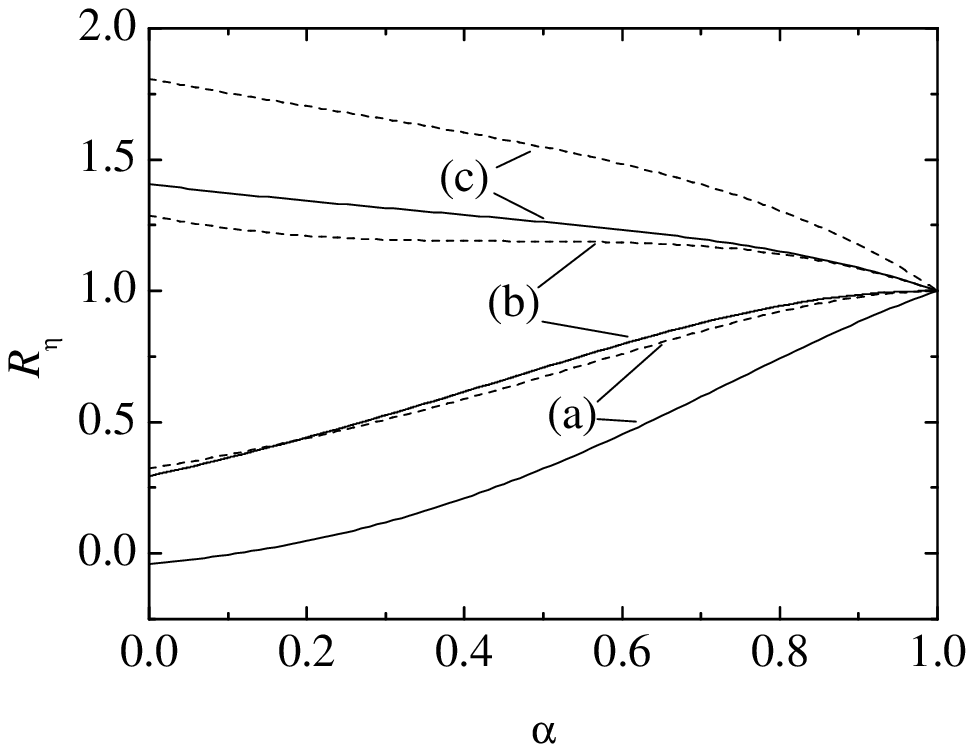}
\caption{Plot of the generalized Prandtl number 
$R_\eta(\alpha)$ from the Boltzmann equation for three-dimensional  IMM 
(solid lines) and IHS (dashed lines) for (a) undriven systems, (b) Gaussian 
thermostat and (c) white noise thermostat. The curves corresponding to IMM in case (b) and to IHS in case (a) are very close each other by accident.\label{fig3}}
\end{figure}
\begin{figure}[tbp]
\includegraphics[width= \columnwidth]{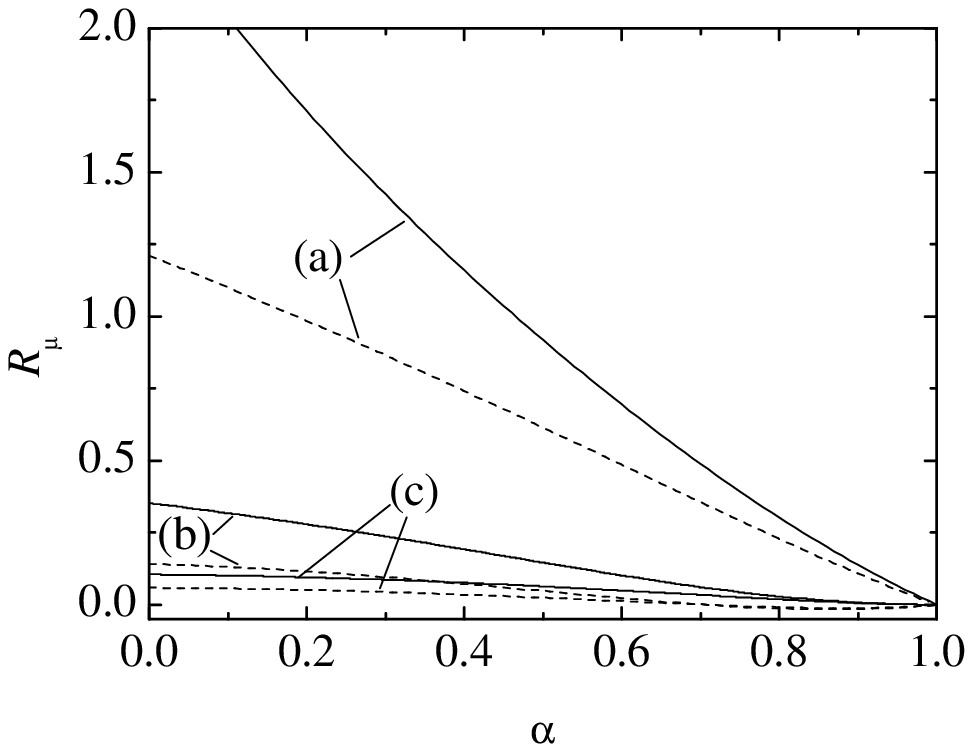}
\caption{Plot of the ratio $R_\mu(\alpha)$ from the Boltzmann equation for 
three-dimensional IMM (solid lines) and IHS (dashed lines) for 
(a) undriven systems, (b) Gaussian thermostat and (c) white noise 
thermostat.\label{fig4}}
\end{figure}

\begin{figure}[tbp]
\includegraphics[width= \columnwidth]{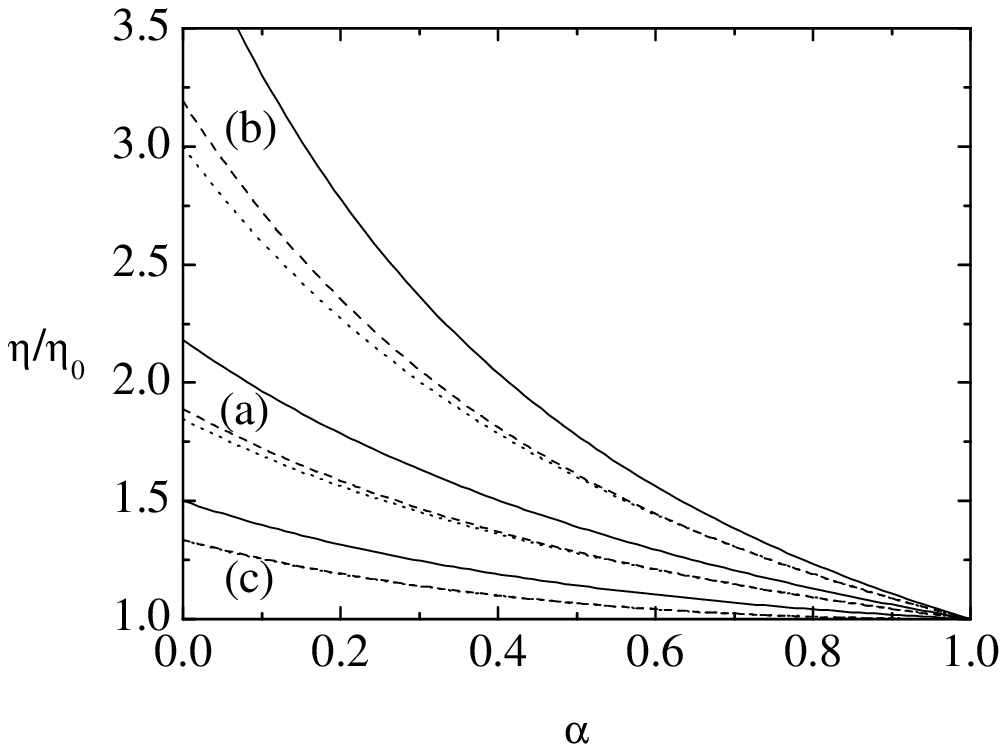}
\caption{Plot of the reduced shear viscosity $\eta/\eta_0$ from the 
Boltzmann equation for three-dimensional  IMM (solid lines) and 
IHS (dashed lines), as well as from the BGK-like model of IHS (dotted lines), 
 for (a) undriven systems, (b) Gaussian thermostat and (c) white noise 
thermostat.
The dashed and dotted lines are practically indistinguishable in case (c).
\label{fig5}}
\end{figure}

\begin{figure}[tbp]
\includegraphics[width= \columnwidth]{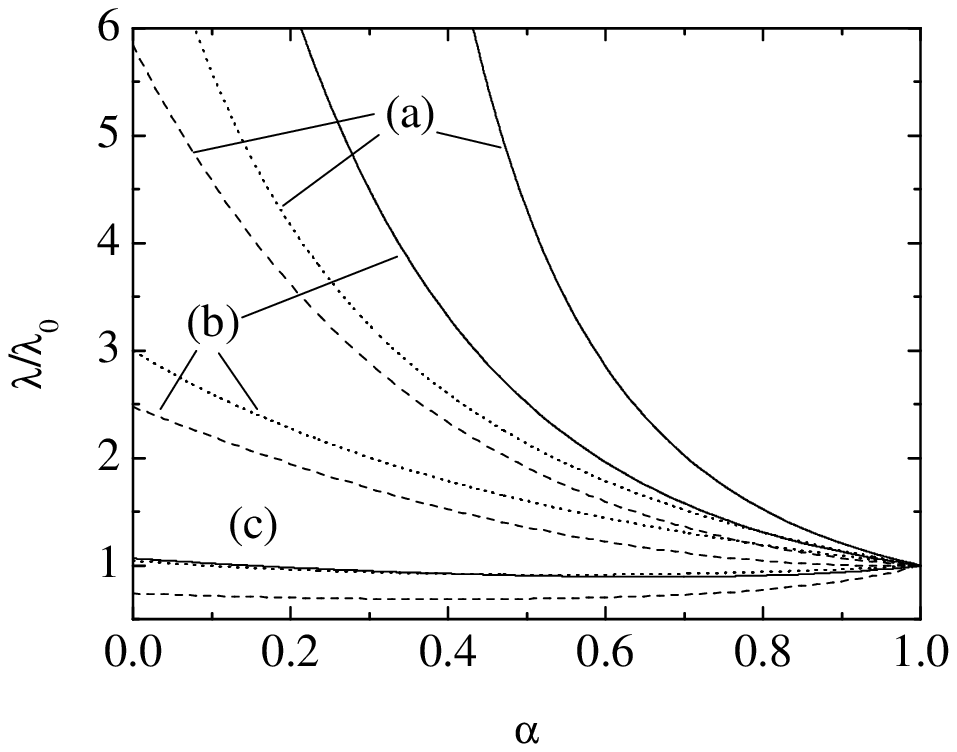}
\caption{Plot of the reduced thermal conductivity  
$\lambda/\lambda_0$ from the Boltzmann equation for three-dimensional IMM 
(solid lines) and IHS (dashed lines), as well as from the BGK-like model of 
IHS (dotted lines), 
 for (a) undriven systems, (b) Gaussian thermostat and (c) white noise 
thermostat. Note that the curves corresponding to IMM and the BGK-like model in case (c) are hardly distinguishable.\label{fig6}}
\end{figure}

\begin{figure}[tbp]
\includegraphics[width= \columnwidth]{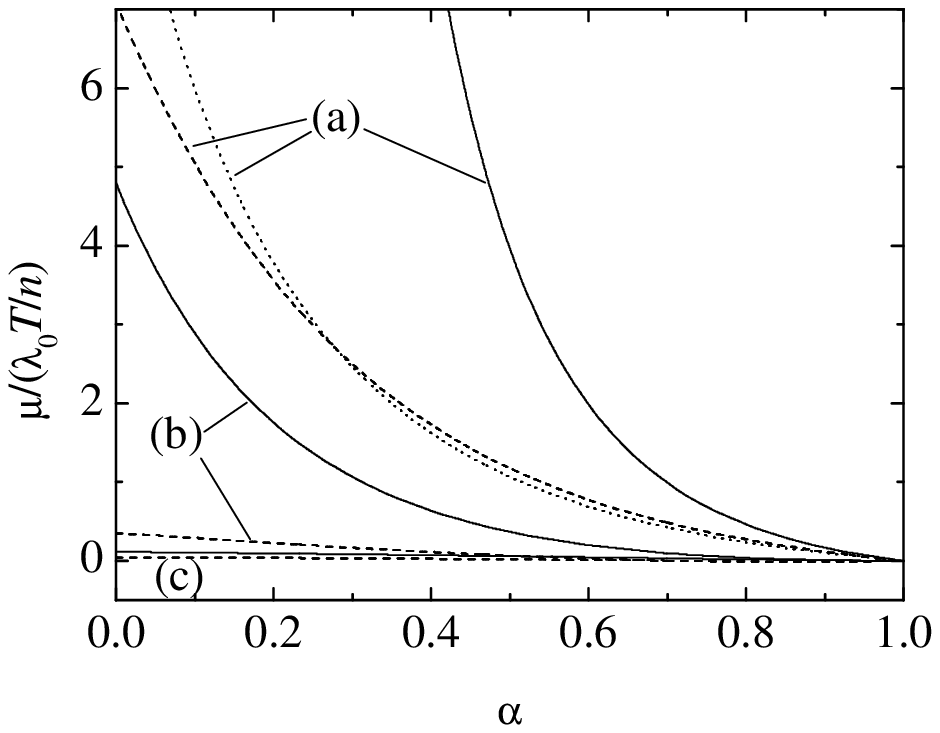}
\caption{Plot of the reduced transport coefficient $\mu/(T\lambda_0/n)$ 
from the Boltzmann equation for three-dimensional  IMM (solid 
lines) and IHS (dashed lines), as well as from the BGK-like model of IHS (dotted line), 
 for (a) undriven systems, (b) Gaussian thermostat and (c) white noise 
thermostat. Note that  
the BGK-like model predicts $\mu=0$ in cases (b) and (c).\label{fig7}}
\end{figure}

Of course, the most interesting comparison refers to the three transport 
coefficients themselves, rather than to their ratios. 
In order to have 
explicit expressions for the transport coefficients of IMM, we now need 
a criterion to fix the free parameter $\omega(\alpha)$. The most natural 
choice to optimize the agreement with the IHS results is to guarantee that 
the cooling rate for IMM, Eq.~(\ref{10}), be the same as that for IHS. 
Strictly speaking, the cooling rate for IHS depends in general on the details of the 
nonequilibrium velocity distribution function, while the effective 
collision frequency $\omega(\alpha)$ in Eq.~(\ref{1}) is assumed to depend 
on $f$ only through the density and the granular temperature. Otherwise, the 
complete knowledge of $\omega$ would require to solve first the Boltzmann 
equation for IHS and then evaluate the cooling rate associated with such a 
solution, what is impractical. Therefore, here I take for $\zeta$ the 
cooling rate of IHS at local equilibrium, namely,
\beq
\zeta=\nu_0 \frac{d+2}{4d}(1-\alpha^2),
\label{e1}
\eeq
where $\nu_0$ is given by Eq.~(\ref{B4}). Making use of Eqs.~(\ref{10}) and (\ref{c11}), 
this is equivalent to
\beq
\omega=\nu_0 \frac{d+2}{2},
\label{e2}
\eeq
\beq
\nu=\nu_0\frac{(1+\alpha)(d+1-\alpha)}{2d}.
\label{e3}
\eeq
With this choice, Eqs.~(\ref{54}), (\ref{63}) and (\ref{c19}) become, respectively,
\beq
\eta=\eta_0 \frac{2d}{1+\alpha}
\left\{
\begin{array}{ll}
4\left[3d+2+(d-2)\alpha\right]^{-1} &\text{(undriven system)},\\
\frac{2}{d}\left(1+\alpha\right)^{-1} &\text{(Gaussian thermostat)},\\
\left(d+1-\alpha\right)^{-1} &\text{(white noise thermostat)},
\end{array}
\right.
\label{e4}
\eeq
\beq
\lambda=\lambda_0 \frac{8(d-1)}{1+\alpha}(1+2a_2)
\left\{
\begin{array}{ll}
\left(d-4+3d\alpha\right)^{-1} &\text{(undriven system)},\\
\frac{1}{2(d-1)}\left(1+\alpha\right)^{-1} &\text{(Gaussian thermostat)},\\
\left[5d+4-(d+8)\alpha\right]^{-1} &\text{(white noise thermostat)},
\end{array}
\right.
\label{e5}
\eeq
\beq
\mu=\frac{T}{n}\lambda_0  \frac{4}{(1+\alpha)^2}
\left\{
\begin{array}{ll}
\frac{2(d+2)(1-\alpha)}{d-4+3d\alpha}+\frac{5d+4-(d+8)\alpha}{d-4+3d\alpha}a_2
&\text{(undriven system)},\\
a_2 &\text{(Gaussian thermostat)},\\
\frac{2(d-1)(1+\alpha)}{5d+4-(d+8)\alpha}a_2 &\text{(white noise thermostat)},
\end{array}
\right.
\eeq
Figures \ref{fig5}--\ref{fig7} compare the three transport coefficients of IMM 
with those of IHS. For completeness, also the coefficients derived from a 
simple BGK-like model (see Appendix \ref{appC}) are included.
Again, the qualitative behaviour of IHS is generally captured by the IMM.
We observe that the shear viscosity increases with the inelasticity, the 
increase being more (less) important when the system is heated with a 
Gaussian (white noise) thermostat. The thermal conductivity increases with 
the inelasticity in a significant way in the undriven case, increases more 
moderately in the case of the Gaussian thermostat, and is almost constant 
in the case of the white noise thermostat. As for the transport coefficient 
$\mu$, it remains small in the thermostatted states, while it 
rapidly increases in the undriven state.
All these trends are, however, strongly exaggerated 
by the IMM in the undriven case (where $\lambda$ and $\mu$ 
diverge at $\alpha=\alpha_0=\frac{1}{9}$) and, to a lesser extent, in the 
Gaussian thermostat case, especially for the coefficient $\mu$. On the other hand, the simple BGK-like model 
\cite{BDS99}
summarized in Appendix \ref{appC} describes fairly well the 
$\alpha$--dependence of the three transport coefficients, being closer to the 
IHS results than the IMM predictions.

\section{Concluding remarks\label{sec6}}
In the Boltzmann equation for inelastic Maxwell models (IMM), the collision rate of the underlying system of inelastic hard spheres (IHS) is replaced by an effective collision rate independent of the relative velocity of the colliding particles. Based on the experience in the case of elastic particles, 
one might reasonably think that this model, while making the collision operator mathematically more tractable, is able to capture the most important properties of IHS, at least those relatively insensitive to the domain of velocities much larger than the thermal velocity.
In particular, one might expect the three transport coefficients (shear viscosity, thermal conductivity and the coefficient associated with the contribution of a density gradient to the heat flux) of IMM to possess a dependence on the  coefficient of restitution similar to that of the transport coefficients of IHS.
The results derived in this paper show, however, that this expectation does not hold true, except at a mild qualitative level, even if the IMM is made to reproduce the cooling rate of IHS. On the other hand, a model kinetic equation based on the well-known BGK model for elastic collisions presents a 
surprisingly good agreement with IHS, once the limitation of the BGK model to reproduce the correct Prandtl number of elastic particles is conveniently accounted for. A partial explanation of this paradox may lie in the fact that in the IMM a rough approximation (collision rate independent of the relative velocity) coexists with the remaining complexity of the detailed collision process. As a result, only a free parameter (essentially the cooling rate) is available to make contact with IHS. In the BGK-like model, however, the cooling rate and its dominant effect on the velocity distribution function is explicitly taken out; what remains of the Boltzmann collision operator is modelled by a conventional relaxation-time term reflecting the effects of collisions not directly associated with the energy dissipation. As a consequence, in addition to the cooling rate, the BGK-like model incorporates an effective (velocity-independent) collision frequency  that increases the flexibility of the model in spite of its simplicity.

In conclusion, the IMM is interesting as a mathematical toy model to explore how a small degree of collisional inelasticity may have a strong influence on the physical properties of the system. For instance, the velocity moments in homogeneous states can be evaluated only approximately in IHS \cite{vNE98}, while they can be exactly obtained in IMM \cite{BCG00}; the high energy tail is another example where rather detailed information can be obtained from the Boltzmann equation for IMM \cite{EB01a,EB01b,KBN01,BMP01,EB01c}. On the other hand, if one is looking for a ``shortcut'' to know some of the properties of IHS, 
the use of the IMM requires a great deal  of caution. 
First, the solution of the Boltzmann equation for IMM may still be a formidable task, except in some special cases. Second, the results derived from the IMM may not be sufficiently representative of the behaviour of IHS, as exemplified here in the case of the transport coefficients.
{}From that point of view, it seems preferable to make use of the BGK-like model proposed in Ref.\ \cite{BDS99}. 
For instance, this model has an exact solution for the nonlinear planar Couette flow (with combined momentum and energy transport) that compares quite well with DSMC data for IHS \cite{TTMGSD01}; for this flow, however, the Boltzmann equation for IMM cannot be solved in a closed form, even in the elastic limit \cite{SG95}.

\ack
I am grateful to Dr.\ J.W. Dufty for insightful discussions about the subject of this paper and to Dr.\ V. Garz\'o for a critical reading of the manuscript.
This work has been partially supported by the Ministerio de Ciencia y Tecnolog\'{\i}a (Spain) through grant No.\ BFM2001-0718.

\appendix
\section{Collisional moments\label{appA}}
Let us consider the general collisional integral of the form
\beq
I[F]\equiv \int\d\mathbf{v}\, F(\mathbf{v})J[\mathbf{v}|f].
\label{AA1}
\eeq
By following standard steps, $I[F]$ can be written as
\beq
I[F]=\frac{\omega}{2n\Omega_d}
 \int \d{\bf v}_1\int \d{\bf v}_2\int \d\widehat{\bm{\sigma}}f({\bf 
v}_1)f({\bf 
v}_2)\left(\widehat{b}-1\right)\left[F(\mathbf{v}_1)+F(\mathbf{v}_2)\right].
\label{AA2}
\eeq
Now we particularize to $F(\mathbf{v})=\mathbf{V}\mathbf{V}$. {}From the 
collision rule (\ref{4})
it follows that
\beq
\left(\widehat{b}-1\right)({\bf V}_1{\bf V}_1+{\bf V}_2{\bf 
V}_2)=\frac{1+\alpha}{2}({\bf 
g}\cdot\widehat{\bm{\sigma}})\left[(1+\alpha)({\bf 
g}\cdot\widehat{\bm{\sigma}})\widehat{\bm{\sigma}}\widehat{\bm{\sigma}}
-{\bf g}\widehat{\bm{\sigma}}-\widehat{\bm{\sigma}}{\bf g}\right].
\label{AA5}
\eeq
To perform the angular integrations we need the results
\beq
\int \d\widehat{\bm{\sigma}}\, ({\bf 
g}\cdot\widehat{\bm{\sigma}})^2\widehat{\bm{\sigma}}\widehat{\bm{\sigma}}= 
B_2 {\bf g}{\bf g}+\frac{B_1-B_2}{d-1}\left(g^2{\sf I}-{\bf 
g}{\bf g}\right),
\label{AA6}
\eeq
\beq
\int \d\widehat{\bm{\sigma}}\, ({\bf 
g}\cdot\widehat{\bm{\sigma}})\widehat{\bm{\sigma}}= B_1 {\bf g},
\label{AA7}
\eeq
where \cite{EB01b}
\beq 
B_n\equiv \int \d\widehat{\bm{\sigma}}\, (\widehat{\bf 
g}\cdot\widehat{\bm{\sigma}})^{2n}=
\Omega_d\pi^{-1/2}\frac{\Gamma(d/2)\Gamma(n+1/2)}{\Gamma(n+d/2)}
\label{AA8}
\eeq
and $\mathsf{I}$ is the $d\times d$ unit tensor.
Therefore,
\beqa
 I[\mathbf{V}\mathbf{V}]&=&-\omega\frac{1+\alpha}{4nd}\int \d{\bf v}_1\int 
 \d{\bf v}_2\,f({\bf 
v}_1)f({\bf v}_2)\left[\frac{1-\alpha}{d}g^2{\sf 
I}\right.\nonumber\\
&&\left.
+2\frac{d+1-\alpha}{d+2}\left({\bf g}{\bf 
g}
-\frac{1}{d}g^2{\sf I}\right)\right]\nonumber\\
 &=& -\omega\frac{1+\alpha}{2md}\left[(1-\alpha)p{\sf 
I}+2\frac{d+1-\alpha}{d+2}
 \left({\sf P}-p{\sf I}\right)\right].
 \label{AA9}
\eeqa
The cooling rate is $\zeta=-(m/dnT)I[V^2]$, so 
taking the trace of Eq.~(\ref{AA9}) we get Eq.~(\ref{10}).

Next we particularize to $F(\mathbf{v})=V^2\mathbf{V}$.
The collision rule gives
\beqa
\left(\widehat{b}-1\right)(V_1^2{\bf V}_1+V_2^2{\bf 
V}_2)&=&\left(\frac{1+\alpha}{2}\right)^2({\bf 
g}\cdot\widehat{\bm{\sigma}})^2 ({\bf V}_1+{\bf V}_2)\cdot({\sf 
I}+2\widehat{\bm{\sigma}}\widehat{\bm{\sigma}})
\nonumber\\
&&
-\frac{1+\alpha}{2}\left({\bf g}\cdot\widehat{\bm{\sigma}})(V_1^2{\sf 
I}+2{\bf V}_1{\bf V}_1\right.\nonumber\\
&&\left.-V_2^2{\sf I}-2{\bf V}_2{\bf 
V}_2\right)\cdot\widehat{\bm{\sigma}}.
\label{A2}
\eeqa
After performing the angular integrations one has
\beqa
I[V^2\mathbf{V}]&=&
\frac{\omega(1+\alpha)}{4nd}\int \d{\bf v}_1\int \d{\bf v}_2\,f({\bf 
v}_1)f({\bf v}_2)
\nonumber\\
&&\times
\left\{\frac{1+\alpha}{2(d+2)}
({\bf V}_1+{\bf V}_2)\cdot[(d+4)g^2{\sf I}+4{\bf g} {\bf g}]\right.
\nonumber\\
&&\left.
-{\bf g}\cdot(V_1^2{\sf I}+2{\bf V}_1{\bf V}_1-V_2^2{\sf I}-2{\bf 
V}_2{\bf V}_2)\right\}
\nonumber\\
&=&-
\frac{2\omega(1+\alpha)(d-1)}{md(d+2)}\left(1+\frac{d+8}{d-1}\frac{1-\alpha}{4} 
\right){\bf q}.
\label{A3}
\eeqa

Finally, let us evaluate the collisional moment $\mu_4$, where $\mu_s$ is 
defined by the last identity in Eq.~(\ref{13}), i.e.,
\beq
\mu_4=-\frac{1}{2\Omega_d} \int \d\widehat{\bm{\sigma}}\int \d{\bf 
c}_1\int \d{\bf c}_2\,\widetilde{f}({\bf c}_1)\widetilde{f}({\bf 
c}_2)\left(\widehat{b}-1\right)(c_1^4+c_2^4).
\label{AA14}
\eeq
From the collision rule one gets \cite{vNE98}
\beqa
\left(\widehat{b}-1\right)(c_1^4+c_2^4)&=&2(1+\alpha)^2
({\bf g}\cdot\widehat{\bm{\sigma}})^2 ({\bf G}\cdot\widehat{\bm{\sigma}})^2
+\frac{1}{8}(1-\alpha^2)^2({\bf g}\cdot\widehat{\bm{\sigma}})^4\nonumber\\
&&
-(1-\alpha^2)({\bf g}\cdot\widehat{\bm{\sigma}})^2 
\left(G^2+\frac{1}{4}g^2\right)\nonumber\\
&&-4(1+\alpha)({\bf g}\cdot\widehat{\bm{\sigma}})({\bf 
G}\cdot\widehat{\bm{\sigma}})({\bf g}\cdot{\bf G}),
\label{AA15}
\eeqa
where $\mathbf{G}\equiv \frac{1}{2}(\mathbf{c}_1+\mathbf{c}_2)$.
Equations (\ref{AA6}) and (\ref{AA7}) give
\beq
\int \d\widehat{\bm{\sigma}}\, ({\bf g}\cdot\widehat{\bm{\sigma}})^2
({\bf G}\cdot\widehat{\bm{\sigma}})^2
= 
\frac{\Omega_d}{d(d+2)}\left[2({\bf g}\cdot{\bf G})^2+g^2G^2\right],
\label{AA16}
\eeq
\beq
\int \d\widehat{\bm{\sigma}}\, ({\bf g}\cdot\widehat{\bm{\sigma}})
({\bf G}\cdot\widehat{\bm{\sigma}})
= \frac{\Omega_d}{d}({\bf g}\cdot{\bf G}) .
\label{AA17}
\eeq
Therefore,
\beqa
\mu_4&=&-\frac{1+\alpha}{2d} \int d{\bf c}_1\int d{\bf 
c}_2\,\widetilde{f}({\bf c}_1)\widetilde{f}({\bf 
c}_2)
\left\{\frac{2(1+\alpha)}{d+2}\left[2({\bf g}\cdot{\bf 
G})^2+g^2G^2\right]\right.
\nonumber\\
&&\left.
+\frac{3(1-\alpha)^2(1+\alpha)}{8(d+2)}g^4-(1-\alpha)g^2\left(G^2+\frac{1}{4}g^2\right)-4({\bf 
g}\cdot{\bf G})^2\right\}
\label{AA18}
\eeqa
Finally, by taking into account that
\beq
({\bf g}\cdot{\bf G})^2=\frac{1}{4}(c_1^4+c_2^4-2c_1^2c_2^2),
\label{AA19}
\eeq
\beq
g^2G^2=\frac{1}{4}[c_1^4+c_2^4+2c_1^2c_2^2-4({\bf c}_1\cdot {\bf c}_2)^2],
\label{AA20}
\eeq
\beq
g^4=c_1^4+c_2^4+2c_1^2 c_2^2+4({\bf c}_1\cdot {\bf c}_2)^2-4(c_1^2+c_2^2)({\bf 
c}_1\cdot {\bf c}_2),
\label{AA21}
\eeq
we get Eq.~(\ref{22}).

\section{Transport coefficients of inelastic hard spheres\label{appB}}
The  derivation of the Navier--Stokes transport coefficients for IHS can be 
found in Refs.~\cite{BDKS98,GD99,BC01,GM01}. In the first Sonine 
approximation, the results are
\beq
\eta=\frac{p}{\nu_0}
\left\{
\begin{array}{ll}
\left({\nu_\eta^*-\frac{1}{2}\zeta^*}\right)^{-1} &\text{(undriven 
system)},\\
\left({\nu_\eta^*-\zeta^*}\right)^{-1} &\text{(Gaussian thermostat)},\\
{\nu_\eta^*}^{-1} &\text{(white noise thermostat)},
\end{array}
\right.
\label{B1}
\eeq
\beq
\lambda=\frac{p}{m\nu_0}\frac{d+2}{2}\left(1+2a_2^{\text{IHS}}\right)
\left\{
\begin{array}{ll}
\left({\nu_\lambda^*-2\zeta^*}\right)^{-1} 
&\text{(undriven system)},\\
\left({\nu_\lambda^*-
\frac{3}{2}\zeta^*}\right)^{-1} 
&\text{(Gaussian thermostat)},\\
{\nu_\lambda^*}^{-1} 
&\text{(white noise thermostat)},
\end{array}
\right.
\label{B2}
\eeq
\beq
\mu=\frac{T}{n}\frac{\lambda}{1+2a_2^{\text{IHS}}}
\left\{
\begin{array}{ll}
\left({\zeta^*+a_2^{\text{IHS}}
\nu_\lambda^*}\right)\left({\nu_\lambda^*-\frac{3}{2}\zeta^*}\right)^{-1} 
&\text{(undriven system)},\\
a_2^{\text{IHS}} &\text{(Gaussian 
thermostat)},\\
a_2^{\text{IHS}} &\text{(white noise 
thermostat)}.
\end{array}
\right.
\label{B3}
\eeq
In these equations \cite{BC01},
\beq
\nu_0=n\sigma^{d-1}(T/m)^{1/2}\frac{4\Omega_d}{\pi^{1/2}(d+2)},
\label{B4}
\eeq
\beq
\zeta^*=\frac{d+2}{4d}(1-\alpha^2)\left(1+\frac{3}{16}a_2^{\text{IHS}}\right),
\label{B5}
\eeq
\beq
\nu_\eta^*=\frac{3}{4d}\left(1-\alpha+\frac{2}{3}d\right)(1+\alpha)\left(1 
-\frac{1}{32}a_2^{\text{IHS}}\right),
\label{B6}
\eeq
\beq
\nu_\lambda^*=\frac{1+\alpha}{d}\left[\frac{d-1}{2}+\frac{3}{16}(d+8)(1-\alpha)+ 
\frac{4+5d-3(4-d)\alpha}{512}a_2^{\text{IHS}}\right].
\label{B7}
\eeq
The expressions for $a_2^{\text{IHS}}$ are given by Eq.~(\ref{b15}) for the 
undriven case and for the Gaussian thermostat and by Eq.~(\ref{b17}) for the 
white noise thermostat.
\section{BGK-like model\label{appC}}
In its simplest version, the BGK-like model proposed in Ref.~\cite{BDS99} 
reads
\beqa
(\partial_t+{\bf v}\cdot\nabla+\mathcal{F})f({\bf r},{\bf 
v};t)&=&-\left(\nu_{\text{BGK}}-\zeta\right)\left[ f({\bf r},{\bf v};t)
-f_0({\bf r},{\bf v};t)\right]\nonumber\\
&&+
\frac{1}{2}\zeta 
\frac{\partial}{\partial {\bf v}}\cdot \mathbf{V} f(\mathbf{r},\mathbf{v};t),
\label{C1}
\eeqa
where $\nu_{\text{BGK}}\propto n T^{1/2}$ is an effective collision 
frequency and $f_0$ is the local equilibrium distribution function.
The BGK-like model (\ref{C1}) can be interpreted as indicating that
a system of inelastic hard spheres behaves \textit{essentially} as a system of 
\textit{elastic} hard spheres with a modified $\alpha$-dependent rate of collisions and
subjected to the action of a ``friction'' force $-\frac{1}{2}m\zeta\mathbf{V}$, 
which accounts for the energy dissipation in an effective way.
Applying the Chapman--Enskog method to Eq.~(\ref{C1}) one obtains 
Eq.~(\ref{47}) with $f^\zero=f_0$ and the replacement
\beq
\mathcal{L}f^\one\to \left(\nu_{\text{BGK}}-\zeta-\frac{1}{2}\zeta 
\frac{\partial}{\partial {\bf V}}\cdot \mathbf{V}\right)f^\one.
\label{C2}
\eeq
It is then straightforward to get the transport coefficients as
\beq
\eta=p
\left\{
\begin{array}{ll}
\left({\nu_{\text{BGK}}-\frac{1}{2}\zeta}\right)^{-1} &\text{(undriven system)},\\
\left({\nu_{\text{BGK}}-\zeta}\right)^{-1} &\text{(Gaussian thermostat)},\\
{\nu_{\text{BGK}}}^{-1} &\text{(white noise thermostat)},
\end{array}
\right.
\label{C3}
\eeq
\beq
\lambda=\frac{p}{m}\frac{d+2}{2}
\left\{
\begin{array}{ll}
\left(\nu_{\text{BGK}}-\frac{3}{2}\zeta\right)^{-1} &\text{(undriven 
system)},\\
\left(\nu_{\text{BGK}}-\zeta\right)^{-1} &\text{(Gaussian 
thermostat)},\\
\left(\nu_{\text{BGK}}+\frac{1}{2}\zeta\right)^{-1} &\text{(white 
noise thermostat)},
\end{array}
\right.
\label{C4}
\eeq
\beq
\mu=\frac{T}{n}\lambda
\left\{
\begin{array}{ll}
{\zeta}\left({\nu_{\text{BGK}}-\zeta}\right)^{-1} &\text{(undriven 
system)},\\
0 &\text{(Gaussian 
thermostat)},\\
0 &\text{(white noise 
thermostat)}.
\end{array}
\right.
\label{C5}
\eeq

Strictly speaking, the cooling rate $\zeta$ is a functional of the velocity distribution
function through Eq.\ (\ref{b12}). In the spirit of the kinetic model (\ref{C1}), the exact cooling rate 
$\zeta$ is approximated by its local equilibrium expression
\beq
\zeta(\alpha)\to\nu_0\frac{d+2}{4d}(1-\alpha^2),
\label{C6}
\eeq
where $\nu_0\propto nT^{1/2}$ is a collision frequency (independent of $\alpha$)
defined by Eq.\ (\ref{B4}).
It remains to fix the $\alpha$--dependence of $\nu_{\text{BGK}}$. 
Comparison between Eqs.\ (\ref{B1}) and (\ref{C3}) suggests the identification
$\nu_{\text{BGK}}\to \nu_0 \nu_\eta^*$ as the simplest choice. Thus, 
\beq
\nu_{\text{BGK}}=\nu_0\frac{3}{4d}\left(1-\alpha+\frac{2}{3}d\right)(1+\alpha),
\label{C7}
\eeq
where we have set $a_2^{\text{IHS}}\to 0$, in consistency with the local equilibrium
approximation (\ref{C6}) for the cooling rate. Inserting Eqs.\ (\ref{C6})  
and (\ref{C7})
into Eqs.\ (\ref{C3})--(\ref{C5}), one gets
\beq
\eta(\alpha)=\eta_0 \frac{4d}{1+\alpha}
\left\{
\begin{array}{ll}
2\left[4+3d-(4-d)\alpha\right]^{-1} &\text{(undriven system)},\\
\left[1+d+(d-1)\alpha\right]^{-1}
 &\text{(Gaussian thermostat)},\\
\left(3+2d-3\alpha\right)^{-1} &\text{(white noise thermostat)},
\end{array}
\right.
\label{C8}
\eeq
\beq
\lambda(\alpha)=\lambda_0 \frac{4d}{1+\alpha}\left\{
\begin{array}{ll}
2\left[d(1+3\alpha)\right]^{-1}
 &\text{(undriven system)},\\
\left[1+d+(d-1)\alpha\right]^{-1} &\text{(Gaussian thermostat)},\\
2\left[8+5d-(8+d)\alpha\right]^{-1}
&\text{(white noise thermostat)},
\end{array}
\right.
\label{C9}
\eeq
\beq
\mu(\alpha)=\frac{T}{n}\lambda_0
\frac{8(d+2)(1-\alpha)}{(1+\alpha)(1+3\alpha)[1+d+(d-1)\alpha]}
\left\{
\begin{array}{ll}
1 &\text{(undriven system)},\\
0 &\text{(Gaussian thermostat)},\\
0 &\text{(white noise thermostat)},
\end{array}
\right.
\label{C10}
\eeq
In the above equations we have taken into account that $\eta_0/m\lambda_0=2/(d+2)$ 
in the BGK model, while the Boltzmann value is $\eta_0/m\lambda_0=2(d-1)/d(d+2)$.
This discrepancy in the Prandtl number is a known feature of the BGK model for elastic collisions and is due to the fact that the model
contains a single relaxation time.

\end{document}